\documentclass[times, twocolumn]{aastex631} 
\usepackage{graphicx, graphics}
\usepackage{CJK}
\usepackage{bbm}
\definecolor{cyan(process)}{rgb}{0.0, 0.72, 0.92}
\usepackage{amsmath, amssymb, bm}

\usepackage[nodayofweek]{datetime}
\usepackage{multirow}
\input{prepcitationformat.txt}
\usepackage{listings}

\usepackage{tikz}
\usetikzlibrary{shapes.geometric, arrows}
\tikzstyle{startstop} = [rectangle, rounded corners, minimum width=1cm, minimum height=0.5cm, text centered, draw=black, fill=red!10, text width = 1cm]
\tikzstyle{start} = [rectangle, rounded corners, minimum width=2cm, minimum height=0.75cm, text centered, draw=black, fill=red!10, text width = 5cm]
\tikzstyle{io} = [trapezium, trapezium left angle=70, trapezium right angle=110, minimum width=0.75cm, minimum height=0.75cm, text centered, draw=black, fill=blue!20]
\tikzstyle{process} = [rectangle, minimum width=3cm, minimum height=1cm, text centered, draw=black, fill=orange!20, text width=6cm]
\tikzstyle{decision} = [diamond, aspect=3, minimum width=0.75cm, minimum height=0.75cm, text centered, draw=black, fill=green!20, text width=4cm]
\tikzstyle{arrow} = [thick,->,>=stealth, rounded corners]

\lstdefinestyle{custom_adql}{
  belowcaptionskip=\baselineskip,
  breaklines=true,
  frame=L,
  xleftmargin=\parindent,
  language=SQL,
  showstringspaces=false,
  basicstyle=\footnotesize\ttfamily,
  keywordstyle=\bfseries\color{green!40!black},
  commentstyle=\itshape\color{purple!40!black},
  identifierstyle=\color{blue},
  stringstyle=\color{orange},
}

\makeatletter
\newcommand*{\centerfloat}{%
  \parindent \z@
  \leftskip \z@ \@plus 1fil \@minus \textwidth
  \rightskip\leftskip
  \parfillskip \z@skip}
\makeatother

\newdateformat{monthyearday}{%
  \THEYEAR\ \monthname[\THEMONTH] \twodigit{\THEDAY}}
\newcommand\numberthis{\addtocounter{equation}{1}\tag{\theequation}}

\newcommand{\Qphi}{$\mathcal{Q}_\phi$}

\newcommand{\uat}[2]{\href{http://vocabs.ands.org.au/repository/api/lda/aas/the-unified-astronomy-thesaurus/current/resource.html?uri=http://astrothesaurus.org/uat/#1}{#2 (#1)}}

\newcommand{\affilCaltechAstro}{\affiliation{Department of Astronomy, California Institute of Technology, MC 249-17, 1200 East California Boulevard, Pasadena, CA 91125, USA}}

\received{2022 May 24}
\revised{2022 October 6} 
\accepted{2022 October 6}
\acceptjournal{The Astrophysical Journal Supplement Series}

\shorttitle{Spiral Flyby Analysis}
\shortauthors{Shuai et al.}

\begin{document}
\pagenumbering{arabic}
\begin{CJK*}{UTF8}{gbsn}
\title{Stellar Flyby Analysis for Spiral Arm Hosts with \textit{Gaia} DR3}

\author[0000-0001-6773-6803]{Linling Shuai (帅琳玲)}
\affiliation{Department of Astronomy, Xiamen University, 1 Zengcuoan West Road, Xiamen, Fujian 361005, China}
\affiliation{Universit\'{e} C\^{o}te d'Azur, Observatoire de la C\^{o}te d'Azur, CNRS, Laboratoire Lagrange, F-06304 Nice, France; \url{bin.ren@oca.eu}}

\author[0000-0003-1698-9696]{Bin B. Ren (任彬)}
\affiliation{Universit\'{e} Grenoble Alpes, CNRS, Institut de Plan\'{e}tologie et d'Astrophysique (IPAG), F-38000 Grenoble, France}
\affiliation{Universit\'{e} C\^{o}te d'Azur, Observatoire de la C\^{o}te d'Azur, CNRS, Laboratoire Lagrange, F-06304 Nice, France; \url{bin.ren@oca.eu}}
\affilCaltechAstro

\author[0000-0001-9290-7846]{Ruobing Dong  (董若冰)}
\affiliation{Department of Physics \& Astronomy, University of Victoria, Victoria, BC, V8P 5C2, Canada}

\author{Xingyu Zhou (周星宇)}
\affiliation{Kavli Institute for Astronomy and Astrophysics, Peking University, Yiheyuan 5, Haidian Qu, 100871 Beijing, People's Republic of China}
\affiliation{Department of Astronomy, Peking University, Yiheyuan 5, Haidian Qu, 100871 Beijing, People's Republic of China}

\author{Laurent Pueyo}
\affiliation{Space Telescope Science Institute (STScI), 3700 San Martin Drive, Baltimore, MD 21218, USA}

\author[0000-0002-4918-0247]{Robert J. De~Rosa}
\affiliation{European Southern Observatory, Alonso de C\'ordova 3107, Vitacura, Santiago, Chile}

\author[0000-0002-2853-3808]{Taotao Fang  (方陶陶)}
\affiliation{Department of Astronomy, Xiamen University, 1 Zengcuoan West Road, Xiamen, Fujian 361005, China}

\author[0000-0002-8895-4735]{Dimitri Mawet}
\affilCaltechAstro
\affiliation{Jet Propulsion Laboratory, California Institute of Technology, 4800 Oak Grove Drive, Pasadena, CA 91109, USA}

\begin{abstract}
Scattered light imaging studies have detected nearly two dozen spiral arm systems in circumstellar disks, yet the formation mechanisms for most of them are still under debate. Although existing studies can use motion measurements to distinguish leading mechanisms such as planet-disk interaction and disk self-gravity, close-in stellar flybys can induce short-lived spirals and even excite arm-driving planets into highly eccentric orbits. With unprecedented stellar location and proper motion measurements from \textit{Gaia}~DR3, here we study for known spiral arm systems their flyby history with their stellar neighbours by formulating an analytical on-sky flyby framework. For stellar neighbors currently located within $10$~pc from the spiral hosts, we restrict the flyby time to be within the past $10^4$~yr and the flyby distance to be within $10$ times the disk extent in scattered light. Among a total of $12570$ neighbors that are identified in \textit{Gaia}~DR3 for $20$ spiral systems, we do not identify credible flyby candidates for isolated systems. Our analysis suggests that close-in recent flyby is not the dominant formation mechanism for isolated spiral systems in scattered light.
\end{abstract}


\section{Introduction}
Existence of substructures in protoplanetary disks, such as gaps, spirals, and cavities, can reveal ongoing planet-disk interaction \citep[e.g.,][]{zhu11, dong15, zhang18, bae19, long22}. Among these planet-disk interaction features, spiral arms could be driven by young giant planets that formulate the best targets for direct imaging observation and characterization, in the sense that these planets are massive and located far from their central stars  \citep[e.g.,][]{dong15, bae16, bae18}. Nevertheless, there have been no confirmed planets that drive spirals \citep[e.g.,][]{brittain20}, making it a necessity to test known spiral arm formation mechanisms on the co-existence of spirals and planets.

For isolated systems, two leading mechanisms can produce spiral arms \citep[][]{dong18}: companion-disk interactions (e.g., stars, brown dwarfs, or giant planets), or gravitational instability in massive disks. Given the distinct rotation patterns under these two scenarios, spiral motion measurements can constrain the locations of companions in the companion-disk interaction scenario \citep[][]{ren18, safonov22}, and even distinguish the two mechanisms \citep{ren20, xie21}, while assuming circular orbits for companions. Based on spiral motion measurements, reported candidates in a couple of systems have been ruled out as the drivers of the observed spiral arms 
\citep[e.g.,][]{ren18, boccaletti21}. 

Close-in stellar flybys can also trigger spiral arms \citep[e.g.,][]{cuello19, menard20}. Spiral pattern formed in this way, however, dissipates on the order of 1000 years after the periastron passage \citep[e.g.,][]{pfalzner03, cuello19}. Despite its short time scale, flyby can leave significant observational effects including disk truncation \citep[e.g.,][]{pfalzner21} and exciting planetary objects to far-separation orbits with high eccentricity \citep[e.g.,][]{derosa19, nguyen21}, complicating the detection of the expected spiral-arm-driving planets. 

\begin{figure*}[htb!]
	\centerfloat
	\includegraphics[width=0.95\textwidth]{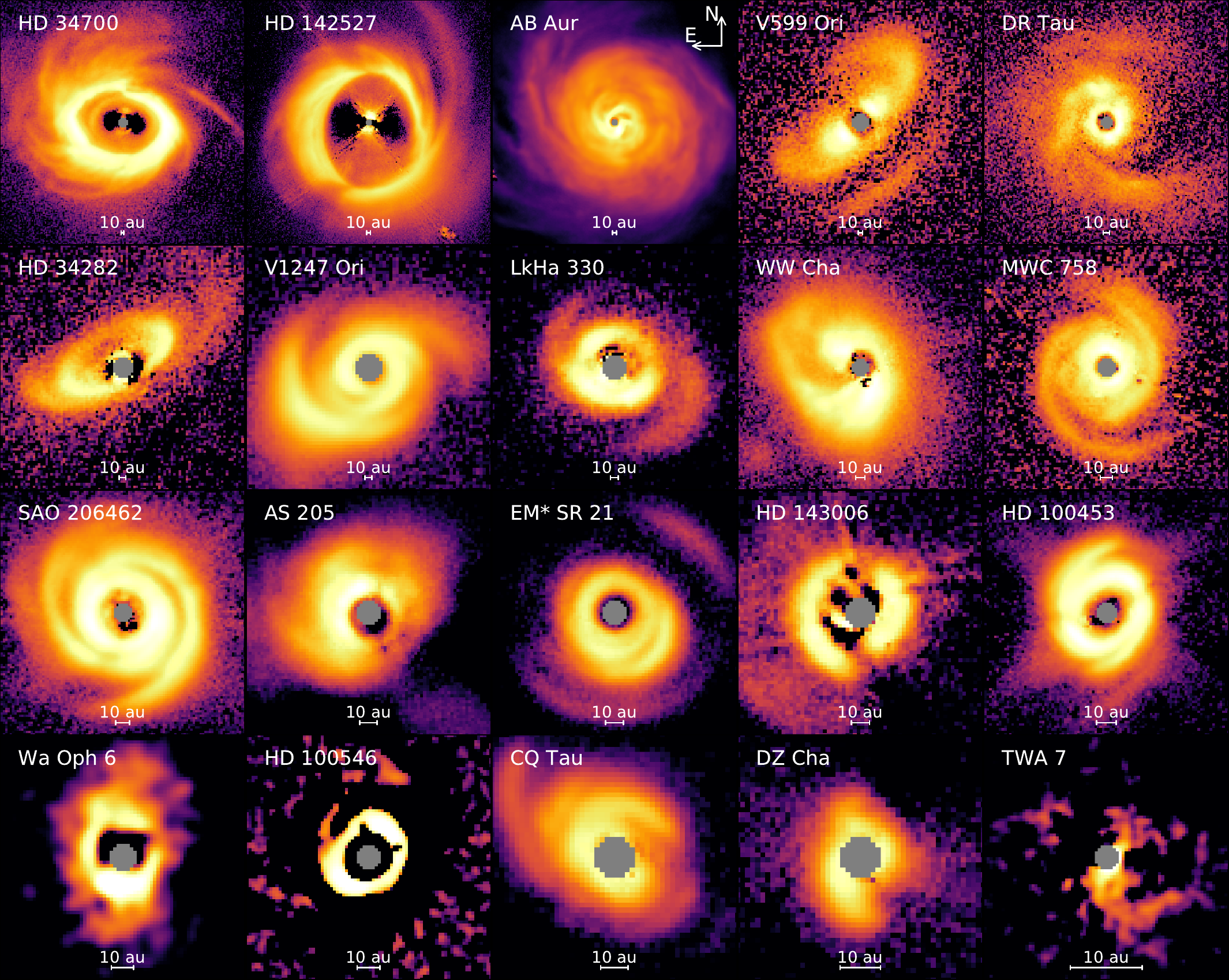}
    \caption{Gallery of spiral systems studied here for flyby analysis (colorbar in log scale). The targets are sorted by decreasing physical extent from left to right, and top to bottom.}
    \label{fig-gallery}
\end{figure*}

With the state-of-the-art high precision measurements on the location and proper motion of stars from \textit{Gaia} DR3 \citep{gaiaDR3}, here we present an investigation into the close-in flyby history of known spiral arm systems. Using Monte Carlo simulations, \citet{derosa19} have studied the stellar flyby history for the HD~106906 circumstellar disk host. They presented a compelling explanation to the large-separation and high-eccentricity of the planet HD~106906~b \citep{nguyen21}. In comparison, analytical multivariate statistical approach has been used to determine association memberships \citep[e.g.,][]{gagne14, riedel17, gagne18}. Here we adopt the latter strategy\footnote{The approach distances in this study are on-sky two-dimensional distances, rather than true three-dimensional distances.} for computational feasibility of ensemble analysis on existing spiral systems.

\textit{Gaia} DR3 products follow multivariate normal distributions \citep{gaiaDR3}, which enables us to develop an analytical framework for flyby analysis. We identify the spiral arm hosts and calculate their flyby history in Section~\ref{sec-dra}, discuss the results in Section~\ref{sec-disc}, and summarize the study in Section~\ref{sec-sum}. We present a querying example using \textit{Gaia}~DR3 in \ref{sec-gaia-query-code}, and the mathematical framework for the flyby analysis in \ref{sec-app-unc}.

\section{Data Reduction and Analysis}\label{sec-dra}

\subsection{Target selection and data retrieval}
We retrieve the public SPHERE/IRDIS observations \citep{dohlen08, vigan10} in polarimetric imaging mode \citep{deboer20, vanholstein20} from the Very Large Telescope (VLT) at the ESO data archive as of 2022 May 24, reduce the observations with {\tt IRDAP} from \citet{vanholstein17, vanholstein20} using the default reduction parameters. From all the reduction results, we identify 20 systems that host spiral arms: AB~Aur, AS~205, CQ~Tau, DR~Tau, DZ~Cha, EM*~SR~21, HD~34282, HD~34700, HD~100453, HD~100546, HD~142527, HD~143006, LkHa~330, MWC~758, SAO~206462 (HD~135344~B), TWA~7, V599~Ori, V1247~Ori, Wa~Oph~6, and WW~Cha. 

We present the stellar-signal-removed \Qphi~images that reveal primarily disk signals in Figure~\ref{fig-gallery} in decreasing physical extent with a colorbar in log scale. To better reveal the spiral structures, we have processed for Wa~Oph~6 by convolving the image with a $\sigma=2$~pixel Gaussian kernel for smoothing, for HD~100546 a high-pass filter by first removing a $\sigma=2$~pixel Gaussian convolution then applying a $\sigma=1$~pixel Gaussian convolution for smoothing, for TWA~7 the disk-removed residual from \citet{ren21}.

\begin{figure}
\centering
\begin{tikzpicture}[node distance=0cm]
\node (start) [start] {Input One Spiral Host};
\node (in1) [io, below of=start, yshift=-1.25cm] {Input One Neighboring Star};
\draw [arrow] (start) -- (in1);
\node (pro1) [process, below of=in1, yshift=-1.25cm] {Calculate Closest Approach};
\draw [arrow] (in1) -- (pro1);
\node (dec1) [decision, below of=pro1, yshift=-2cm] {Closest Approach Distance $d_{\rm c.a.} \in [0, 10\, r_{\rm disk}]$?};
\draw [arrow] (pro1) -- (dec1);
\node (dec2) [decision, below of=dec1, yshift=-2.75cm] {Approach within $10\, r_{\rm disk}$ at $t \in [-10^4, 0]$~yr?};
\draw [arrow] (dec1) -- node[anchor=east] {yes} (dec2);
\node (pro3) [process, below of=dec2, yshift=-2.25cm] {Inspect Relative Distance with Time};
\draw [arrow] (dec2) -- node[anchor=east] {yes} (pro3);
\node (dec3) [decision, below of=pro3, yshift=-2.25cm] {Uncertainty Grows Faster than Distance?};
\draw [arrow] (pro3) -- (dec3);
\node (pro4) [process, below of=dec3, yshift=-2.25cm] {MCMC Quantification of Closest Approach Time};
\draw [arrow] (dec3) -- node[anchor=east] {no} (pro4);
\node (pro5) [process, below of=pro4, yshift=-1.5cm] {Monte Carlo Quantification of Closest Approach Distance};
\draw [arrow] (pro4) -- node[anchor=east] {} (pro5);
\node (stop4) [startstop, below of=pro5, yshift=-1cm] {Stop};
\draw [arrow] (pro5) -- node[anchor=south] {} (stop4);
\draw [arrow] (dec1) -- +(4.25, 0) node[below, midway] {no} |-  (stop4);
\draw [arrow] (dec2) -- +(4.25, 0) node[below, midway] {no} |-  (stop4);
\draw [arrow] (dec3) -- +(4.25, 0) node[below, midway] {yes} |-  (stop4);
\end{tikzpicture}
\caption{Flowchart of the flyby framework. For a pair of stars, three conditions have to be satisfied for MCMC quantification of the time and distance of the closest approach.}\label{fig-flowchart}
\end{figure}
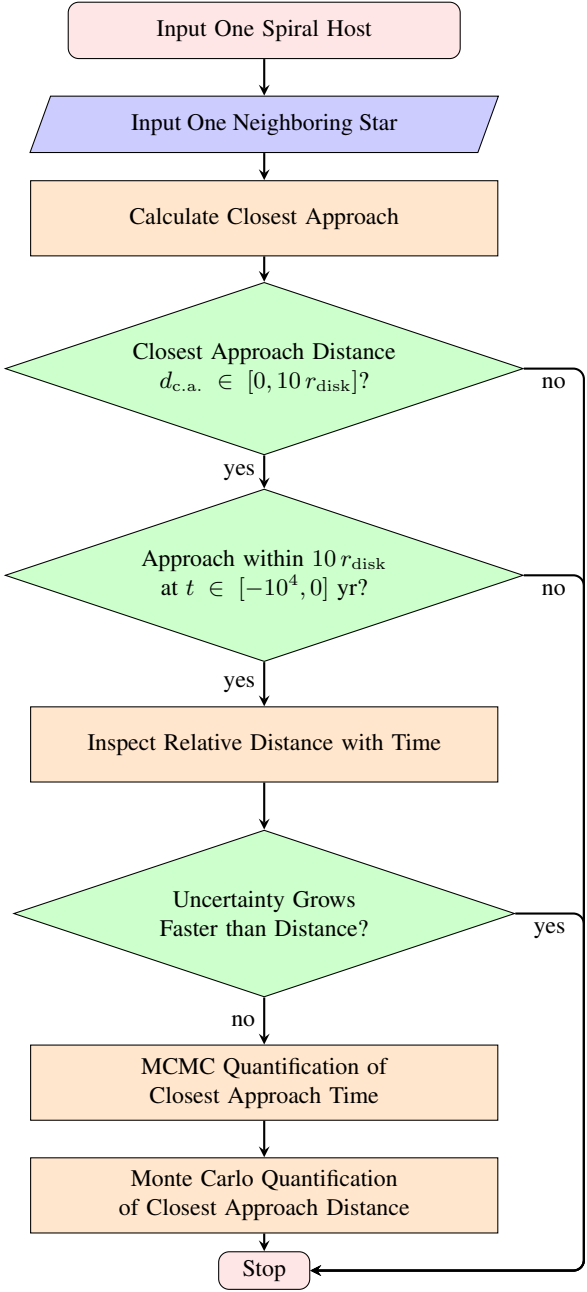

To investigate recent close-in flyby history for each spiral arm host, we query its neighboring stars within $10$~pc from the \textit{Gaia} DR3 database \citep{gaiaDR3} for further analysis, see \ref{sec-gaia-query-code} for an example. 

Given that \textit{Gaia} DR3 provides only location and proper motion measurements in both right ascension ($R.A.$) and declination ($Decl.$) directions, and it does not offer radial velocity measurements for all stars, we do not attempt to perform three dimensional kinematic studies \citep[e.g.,][]{derosa19, ma22}. However, our framework can be generalized to take into account of new measurements, see \ref{sec-app-unc}.

\subsection{Flyby framework}
We briefly describe the key concepts of our flyby framework here, see the mathematical details and corresponding derivation in \ref{sec-app-unc}. In this paper, we use $\mathcal{N}_k$, where $k\in\mathbb{N}^+$, to denote a $k$-dimensional multivariate normal distribution,\footnote{Unless otherwise specified, the uncertainties in this study are $1\sigma$ in frequentist probability expressions, or (16th, 84th) percentiles in Bayesian statistics.} and superscript $^T$ for matrix transpose. We develop the analytical flyby framework for spirals ({\tt afm-spirals}; \citealp{shuai22}) for this purpose.\footnote{\url{https://github.com/slinling/afm-spirals}}

\subsubsection{Mathematical construction}
For the selected spiral hosts and their neighboring stars, \textit{Gaia} DR3 products do not have complete radial velocity information for three dimensional motion analysis, we therefore study only two dimensional on-sky location for the selected samples. 

For each star, its on-sky location ($R.A.$~and $Decl.$) and its two dimension proper motion vector, $\bm{q} = (R.A., Decl., {\rm pm}_{R.A.}, {\rm pm}_{Decl.})^T\in\mathbb{R}^{4\times1}$, can follow a four-dimensional multivariate normal distribution, \begin{equation}
    \bm{q}\sim\mathcal{N}_4(\bm{\mu}, \bm{\Sigma}),
\end{equation}
where $\bm{\mu} \in \mathbb{R}^{4\times1}$ and $\bm{\Sigma}\in\mathbb{R}^{4\times4}$ are the expectation and covariance matrices that can be computed using \textit{Gaia}~DR3. With these information, we can obtain the on-sky location for a star at a given time $t\in\mathbb{R}$ by left-multiplying $\bm{q}$ a transformation matrix \[
T_t = \begin{bmatrix}
1& 0 & t & 0\\ 
0 & 1 & 0 & t
\end{bmatrix},
\] and reach a bivariate normal distribution
\begin{equation}
    \bm{x}_t\sim\mathcal{N}_2 (T_t \bm{\mu}, T_t \bm{\Sigma} T_t^T),
\end{equation}
where $\bm{x}_t=(R.A., Decl.)^T\in \mathbb{R}^{2\times1}$ is the on-sky location vector at time $t$.

For a pair of stars, their relative on-sky position is a subtraction of two bivariate normal distributions, which results into a bivariate normal distribution. Therefore, in the reference frame of a spiral host $H$, the relative location of a star neighbour $N$ at any given time $t$, follows \begin{equation}
    \Delta\bm{x}_{NHt}\sim \mathcal{N}_2\left(T_t \bm{\mu}^{(N)}-T_t \bm{\mu}^{(H)}, T_t \bm{\Sigma}^{(N)}T_t^T +T_t\bm{\Sigma}^{(H)} T_t^T \right),
\end{equation}
where superscripts $^{(H)}$ and $^{(N)}$ denote the values for the spiral host and a neighboring star, respectively.

At a certain time $t$, to quantify the statistical significance between spiral host $H$ and its neighboring star $N$, we should calculate the relative distance and statistical distribution along line $\overleftrightarrow{HN}$. Although this distribution can be obtained by marginalizing the bivariate normal distribution along the perpendicular direction of $\overleftrightarrow{HN}$, we ease this mathematical complexity by instead rotating the relative distribution of $N$, thus aligning it along the new horizontal axis using a rotation matrix. With this rotation, the distribution along the direction of $\overleftrightarrow{HN}$ is a 1-dimensional normal distribution, \begin{equation}\label{eq-main-parallel}
x_\parallel \sim \mathcal{N}_1\left(\mu_\parallel, \sigma^2_\parallel \right),\end{equation} where subscript $_\parallel$ denotes the distribution along line $\overleftrightarrow{HN}$. The relative distribution of a star pair at time $t$ is now established, see Appendix~\ref{sec-app-stat-distribution} for the derivation and detailed expression. By minimizing the expected separation in Equation~\eqref{eq-main-parallel}, we can obtain the closest separation using a \textit{frequentist} approach. 

We obtain the closest approach time using a \textit{Bayesian} approach\footnote{We use superscripts $^{(f)}$ and $^{(B)}$ to distinguish the values from frequentist and Bayesian approach, respectively.}, as well as the corresponding closest separation, as follows. For a host-neighbor pair, with the one-dimensional relative distribution between the two stars at time $t$ in Equation~\eqref{eq-main-parallel}, we can obtain the corresponding probability density function. In order to obtain the posterior distribution of the closest approach time $t_{\rm c.a}^{(B)}$ from a Bayesian approach, we can rewrite the probability density function into a log-likelihood expression, 
\begin{equation}\label{eq-main-likelihood}
\ln \mathcal{L} (t \mid H, N) = -\frac{1}{2} \left(\frac{x_\parallel - \mu_\parallel}{\sigma_\parallel} \right)^2 - \ln \sigma_\parallel - \frac{1}{2}\ln(2\pi),
\end{equation}
and perform Markov Chain Monte Carlo (MCMC) analysis. To obtain the distribution for the corresponding distance $d_{\rm c.a.}^{(B)}$, we can draw samples from the posterior distribution of $t_{\rm c.a.}^{(B)}$ and then sample from their corresponding distances, see Appendix~\ref{sec-app-posterior} for the detailed procedure.

\begin{figure*}[hbt!]
	\centerfloat
	\includegraphics[width=0.95\textwidth]{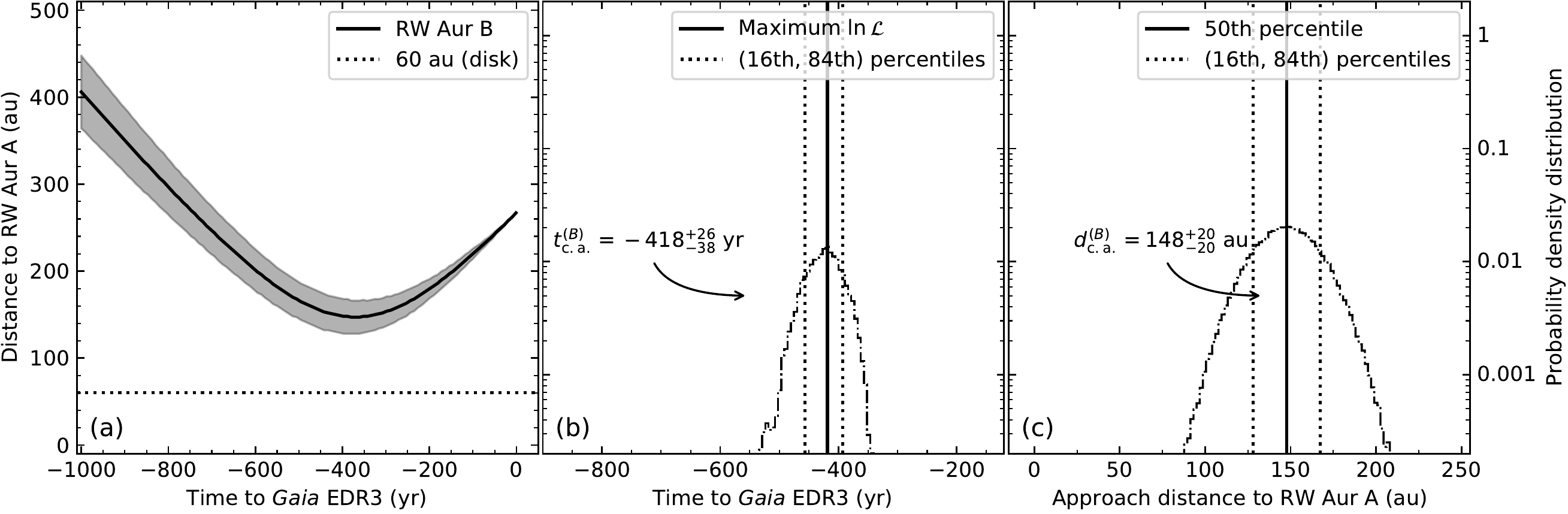}
    \caption{Flyby of RW~Aur~B from RW~Aur~A with \textit{Gaia}~DR3. ({a}) Relative distance between the binary in the past 1000~yr. ({b}) and ({c}) are the posterior distributions of the on-sky closest approach time and separation, respectively. The closest approach is  $d^{(B)}_{\rm c.a.}=148_{-20}^{+20}$~au at $t^{(B)}_{\rm c.a.}=-418_{-38}^{+26}$~yr.}
    \label{fig-rw-aur}
\end{figure*}

\subsubsection{Framework application}
Stellar flyby can excite spiral arms \citep[e.g.,][]{pfalzner03, menard20}, however there are three requirements for the spirals to be observable: 

(1) the flyby should be a relatively recent or even ongoing event, since spirals dissipate after ${\sim}1000$ years  \citep[e.g.,][]{cuello19},

(2) the periastron has to be relatively close to the disk, and 

(3) the flyby needs to occur slow enough to allow gravitational interactions to take place; a parabolic approach is ideal as it induces the greatest angular momentum transfer and the most prominent spiral \citep{pfalzner04, vincke16, cuello19}. 

In order to find potential flybys responsible for the spiral arms, we adopt the following procedure.

For a given spiral arm host, we first calculate the closest historical approach for all its surrounding stars within $10$~pc. Given the fact that the locations are linearly propagated, with the linear approximation already validated in \citet{derosa19}, we can analytically solve for the closest approach distance and time since this calculates the intersection of two lines, see Appendix~\ref{sec-app-ca}. 

If the closest approach distance $d^{(f)}_{\rm c.a.}$ is within 10 times the disk radius, we inspect the distance of the two stars within the past $10^4$~yr. We then explore the uncertainty of the closest approach, if its corresponding closest time $t^{(B)}_{\rm c.a.}$ is within this period.  We note that even if the closest approach occurred more than $10^4$ years ago,
the relative distance of the host-neighbor can still be within 10 times of the disk radius. In our calculation, we have taken into account this scenario. See Figure~\ref{fig-flowchart} for a flowchart of the procedure in this study.

In our experiments, we have noticed that uncertainty could grow faster than distance for some pair of stars (e.g., Section~\ref{sec-hd34700}), which results into higher mathematical likelihoods for farther separations, since the first term in Equation~\eqref{eq-main-likelihood}  can increase to nearly $0$ in under this scenario. We therefore only apply the MCMC framework to those that can be well-characterized with a maximum likelihood approach.

\subsubsection{Framework validation: RW~Aur}
RW~Aur is a binary system in which a ${\sim}600$~au long arm trails from the primary in millimeter observations \citep[][]{cabrit06, rodriguez18}. By reproducing the observed features using hydrodynamic models, the study by \citet{dai15} supported the hypothesis of a tidal encounter between the two stars. Although observations with SPHERE/IRDIS in polarized light from existing ESO programs including 0102.C-0656(A) and 0104.C-0122(A) do not show similar features, it does not preclude the existence of faint structures that are beyond the sensitivity of SPHERE. To explore the capability of our analytical framework, we investigate the flyby history between RW~Aur~A and RW~Aur~B while assuming a linear motion using \textit{Gaia}~DR3 (which could be less accurate than the parabolic flyby study as in \citealp{cuello20}) for demonstration purposes.

Using the location and proper motion measurements from \textit{Gaia}~DR3, we first minimize the expectation for the closest approach distance in Equation~\eqref{eq-main-parallel}, see Figure~\ref{fig-rw-aur}a. We obtain a minimum on-sky separation of $d^{(f)}_{\rm c.a.}=148\pm 20$~au from RW~Aur~A, which is equivalently $7\sigma$ from zero distance, at $t^{(f)}_{\rm c.a.}=-371$~yr. We then explore the likelihood in Equation~\eqref{eq-main-likelihood} using {\tt emcee}, and obtain the posterior distribution of the closest approach time in Figure~\ref{fig-rw-aur}b, or $t^{(B)}_{\rm c.a.} = -418_{-38}^{+26}$~yr. We finally sample from the posterior distribution of the approach time, then sample the corresponding approach distance (see the prescription in Appendix~\ref{sec-app-posterior-distance}), to reach the corresponding posterior distribution for the closest approach distance in Figure~\ref{fig-rw-aur}c, or $d^{(B)}_{\rm c.a.}=148_{-20}^{+20}$~au. 

The flyby between the RW~Aur binary has demonstrated the applicability of our framework. In addition, the orbital period of the binary is more than $1500$~yr \citep{bisikalo12} and the orbit could be parabolic \citep{dai15}, making it likely that the flyby only occurs within a small fraction of the orbit and thus a linear propagation is valid. We note, however, that our flyby distance and time for the flyby of the RW~Aur system could be more accurate when more realistic orbits are adopted \citep[e.g.,][]{cuello20}.

In fact, the \textit{Gaia}~DR3 measurements for RW~Aur may suffer from systematics that can lead to incorrect flybys in our study for RW~Aur. Specifically, the renormalized unit weight error (RUWE) is 1.50 for RW~Aur~A, while RUWE = 16.4 for RW~Aur~B. In comparison with an expected value of 1.0 for the RUWE, the values for the RW~Aur binary are higher than 1.4, which suggests unreliable \textit{Gaia}~DR3 solutions \citep[e.g.,][]{fabricius21}. Given that the closeness of the two stars may lead to inaccurate measurements \citep[e.g.,][]{takami20}, future high precision measurements with better \textit{Gaia} solutions for non-isolated systems, as well as radial velocity measurements, could better determine the flyby history for RW~Aur.

Out of the 12130 neighbors that are within $30$~pc from RW~Aur, we additionally identify that \textit{Gaia}~DR3~156431440590447744 has approached $d^{(f)}_{\rm c.a.} =13.2\pm 2.5$~kau from RW~Aur at $t^{(f)}_{\rm c.a.}=-5845$~yr. With large uncertainties, we do not perform Bayesian analysis for this neighbor using Equation~\eqref{eq-main-likelihood}. However, it is 5$\sigma$ from zero on-sky distance with RW~Aur~A and might have flown by statistically. In comparison, it is 7$\sigma$ for RW~Aur~B, and the rest of the neighbors are at least $300\sigma$ away in their closest approach.

\begin{deluxetable}{l rcc rrr c}[htb!]
\setlength{\tabcolsep}{3pt}
\tablecaption{Summary of flyby information for 20 spiral hosts\label{tab:flybyinfo}}
\tablehead{
Star	&	$r_{\rm disk}$	&	$n_{\rm n.b.r}$	&	$n_{\rm flyby}$	&	$d_{\rm c.a.}^{(f)}$\tablenotemark{a}	&	$\sigma_{d_{\rm c.a.}}^{(f)}$	&	$z_{\rm min}$\tablenotemark{a}	&	RUWE  \\
           	&	(au)	&	 	&	 	&	(au)	&	(au)	&	 	&	 \\
	(1)	&	(2)	&	(3)	&	(4)	&	(5)	&	(6)	&	(7)	&	(8) }
\startdata
EM*~SR~21       	&	68	&	1432	&	13	&	203	&	133	&	1.52$\sigma$		&	1.14		\\
HD~34700    	&	438	&	367	&	7	&	3184	&	127	&	25$\sigma$		&	1.05		\\ \hline
AB~Aur      	&	312	&	492	&	7	&	339445	&	130	&	2604$\sigma$		&	1.37		\\
CQ~Tau      	&	45	&	406	&	9	&	112693	&	15	&	316$\sigma$		&	3.70\tablenotemark{b}	\\
DZ~Cha      	&	30	&	378	&	8	&	144168	&	48	&	1128$\sigma$		&	1.51\tablenotemark{b}		\\
HD~34282    	&	185	&	327	&	12	&	20003	&	24	&	726$\sigma$		&	1.41\tablenotemark{b}		\\
HD~100546   	&	54	&	654	&	10	&	115309	&	134	&	512$\sigma$		&	1.30	\\
HD~142527   	&	318	&	1973	&	28	&	56125	&	117	&	37$\sigma$		&	1.18		\\
HD~143006   	&	67	&	525	&	8	&	19620	&	133	&	147$\sigma$		&	1.13		\\
LkHa~330    	&	159	&	537	&	4	&	443374	&	3	&	1136$\sigma$		&	1.85\tablenotemark{b}		\\
MWC~758     	&	101	&	458	&	5	&	39479	&	198	&	199$\sigma$		&	0.99		\\
SAO~206462  	&	88	&	710	&	11	&	55053	&	291	&	189$\sigma$		&	0.91		\\
TWA~7       	&	17	&	201	&	6	&	57813	&	0	&	14326$\sigma$		&	1.19		\\
V1247~Ori   	&	180	&	606	&	2	&	945081	&	646	&	1463$\sigma$		&	1.25		\\
Wa~Oph~6    	&	55	&	529	&	15	&	18587	&	34	&	30$\sigma$		&	1.25		\\
WW~Cha      	&	132	&	558	&	16	&	3061	&	173	&	18$\sigma$		&	2.80\tablenotemark{b}	\\
DR~Tau      	&	193	&	422	&	9	&	115191	&	2892	&	40$\sigma$		&	1.87\tablenotemark{b}		\\ \hline
AS~205      	&	71	&	793	&	6	&	214026	&	2484	&	86$\sigma$		&	1.75\tablenotemark{b}		\\
HD~100453   	&	62	&	816	&	10	&	1254	&	12	&	2$\sigma$		&	1.08		\\
V599~Ori    	&	282	&	386	&	3	&	501369	&	86	&	5769$\sigma$		&	1.30		\\
\enddata
\tablecomments{
Column 1: spiral host name. Column 2: approximate disk radius in scattered light matching field of view in Figure~\ref{fig-gallery}. Column 3: number of neighboring stars in \textit{Gaia}~DR3 with location and proper motion measurements within 10~pc. Column 4: number of neighboring star whose closest approach to the host are within $10^4$~yr before \textit{Gaia}~DR3. Columns 5 and 6: expectation and standard deviation for closest on-sky approach distance in the frequentist approach. Column 7: minimum $z$-score for the closest approaches of all neighbors, which is defined as the expectation of closest approach by its corresponding standard deviation; the resulting value is thus unitless or the meaning of $\sigma$ in astronomy. Column 8:
Renormalised Unit Weight Error (RUWE) of the host star in \textit{Gaia}~DR3.
\tablenotetext{a}{Closest approach distance and minimum $z$-score do not have to be the same neighbor. If not, the $z$-score for the closest approach is higher, which is even less likely to pass through the spiral host.}
{\tablenotetext{b}{The RUWE is expected to be around 1.0 for sources where the single-star model provides a good fit to the astrometric observations. An RUWE value that is significantly greater than 1.0 (e.g., RUWE ${>}1.4$, \citealp{fabricius21}) suggests that the source is non-single or otherwise problematic for the astrometric solution, which would induce biased flyby calculations in our framework. }
}
}
\end{deluxetable}

\section{Results and Discussion}\label{sec-disc}
We report the closest approach results from the frequentist approach in Table~\ref{tab:flybyinfo} for all spiral hosts. For each host, the flyby information is available for all its neighbors in electronic {\tt .csv} tables, see Table~\ref{tab-emsr21-example} for an incomplete example for EM*~SR~21~A.

For all systems, we discuss the flyby history in three categories: those with credible \textit{Gaia}~DR3 flyby candidates (i.e., passing all criteria in Figure~\ref{fig-flowchart}) in Section~\ref{sec-flyby-with-credible}, those without in Section~\ref{sec-flyby-no-credible} and within the latter the possible binary systems in Section~\ref{sec-flyby-binary}. We note that those spiral hosts that do not have credible flyby candidates may change when more accurate location and proper motion solutions are available.

\subsection{With credible \textit{Gaia}~DR3 flyby candidates}\label{sec-flyby-with-credible}
We have identified credible \textit{Gaia}~DR3 flyby candidates for the following 2 systems, see Figure~\ref{fig-dist-year} for corresponding flyby distance. 

\subsubsection{EM*~SR~21 (i.e., EM*~SR~21~A)}
In scattered light, EM*~SR~21 hosts an asymmetric outer ring (${\sim}70$~au), two bright spiral arms inside this ring, and an inner ring detected by VLT/SPHERE \citep{muroarena20}. ALMA observations of thermal dust emission in multiple bands show a cavity which colocates with the spirals \citep{muroarena20}. In addition to the major spirals, \citet{muroarena20} detect a faint kinked spiral between the bright inner spirals and the outer ring, which is consistent with hydrodynamical models of planet-disk interactions and could constrains the location of a possible planet. Using the pyramid wavefront sensor from Keck/NIRC2 in $L'$ band, \citet{uyama20a} reported three marginal point sources that coincide with the spirals.

Among the 1432 \textit{Gaia}~DR3 sources within $10$~pc from EM*~SR~21 (i.e., EM*~SR~21~A), 13 sources have their closest  on-sky approach within the past $10^4$~yr, among which \textit{Gaia}~DR3~6049152203067495808 (i.e., EM*~SR~21~B) had an on-sky distance $d^{(f)}_{\rm c.a.}=203_{-133}^{+133}$~au (equivalent to being $1.52\sigma$ away from zero separation) from the host at $t^{(f)}_{\rm c.a.}=-3873$~yr. The rest of the 12 systems are at least $100\sigma$ away from the spiral host, see Table~\ref{tab-emsr21-example}.

For EM*~SR~21~B, we follow Appendix~\ref{sec-app-posterior-time} and quantify its posterior distribution of closest approach time. The maximum likelihood approach time and (16th, 84th) percentile values are $t^{(B)}_{\rm c.a.}=-4086_{-1010}^{+291}$~yr from \textit{Gaia}~DR3.

We obtain the posterior distribution of approach distance by first sampling from the posterior distribution of closest approach time, then sampling the approach distance using Equation~\eqref{eq-parallel}, see Appendix~\ref{sec-app-posterior-distance} for the detailed procedure. The corresponding maximum likelihood distance and (16th, 84th) percentile values are $d^{(B)}_{\rm c.a.}=205_{-132}^{+132}$~au, see Figure~\ref{fig-dist-year}c for the probability density distribution. 

In fact, EM*~SR~21~A and EM*~SR~21~B have been classified as a binary system \citep{barsony03}. However, they do not appear to be coeval in \citet{prato03}, 
marking them as a false binary from chance alignment based on characterizing their spectral type and luminosity. However, the ages adopted in \citet{prato03} can be affected by the fact that EM*~SR~21~A is actively accreting \citep{fang17}, and late-type stars like EM*~SR~21~B can manifest much younger ages in comparison with model-derived ages \citep{Malo2014}.

To further assess whether the two are gravitationally bounded, we assume the same parallax for the two and treat them as a two-body system. Adopting a stellar mass of $1.66M_{\odot}$ and $0.25M_{\odot}$ for A and B, respectively, from \citet{Herczeg2014}, and using \textit{Gaia}~DR3 parallax and proper motion, the two will be gravitationally unbound if their radial velocity differs by ${\sim}1.7$ km~s$^{-1}$. What is more, the bounded orbital period would be ${\sim}3\times10^4$~yr, which makes the closest approach time cover only ${\sim}10\%$ or the orbit and thus suggests that our linear motion approximation is likely valid. Nevertheless, different measurements on spectral types and thus masses of the two \citep[e.g.,][]{suarez06, furlan09} could result into different age and bounding results. After all, although our framework 
identifies EM*~SR~21~B as a credible on-sky flyby candidate for EM*~SR~21~A, the fact that two are probably bounded suggests the spirals may have formed through companion-disk interaction \citep[e.g.,][]{dong16}. With \textit{Gaia}~DR3 RUWE measurements being smaller than the $1.4$ threshold (i.e., $1.14$ and $1.20$ for EM*~SR~21~A and EM*~SR~21~B, respectively), we do not expect our results to deviate significantly from future high accuracy measurements.

\begin{deluxetable*}{crr rrr rrr}
\setlength{\tabcolsep}{9pt}
\tablecaption{Flyby history of EM*~SR~21 neighbors using \textit{Gaia}~DR3\tablenotemark{a}}\label{tab-emsr21-example}
\tablehead{
\colhead{Designation\tablenotemark{b}} & \colhead{$d^{(f)}_{\rm c.a.}$\tablenotemark{c}} & \colhead{$\sigma_{d^{(f)}_{\rm c.a.}}$\tablenotemark{c}} & \colhead{$t^{(f)}_{\rm c.a.}$\tablenotemark{c}} & \colhead{$z$-score} & \colhead{$R.A.$} & \colhead{$Decl.$} & \colhead{RUWE} \\
\colhead{} & \colhead{(au)} & \colhead{(au)} & \colhead{(yr)} & \colhead{($\sigma$)} & \colhead{($^\circ$)} & \colhead{($^\circ$)}\\ 
\colhead{(1)}	& \colhead{(2)} & \colhead{(3)} & \colhead{(4)} &  \colhead{(5)}  & \colhead{(6)} & \colhead{(7)}& \colhead{(8)} 
}
\startdata
Gaia DR3 6049152203067495808 & 203.03 & 133.42 & -3873 & 1.52 & 246.7929 & -24.3220 & 1.20\\
Gaia DR3 6049158177365995904 & 10380.06 & 29.50 & -1116 & 351.82 & 246.8071 & -24.3048 & 1.05\\
Gaia DR3 6049152130052048640 & 12786.11 & 127.33 & -2714 & 100.42 & 246.7715 & -24.3356 & 1.08\\
Gaia DR3 6049164327759169536 & 20173.36 & 541.90 & -10000\tablenotemark{d}  & 37.23 & 246.7630 & -24.2770 & 1.13\\
$\cdots$ & $\cdots$ & $\cdots$ & $\cdots$ & $\cdots$ & $\cdots$ & $\cdots$ & $\cdots$\\
Gaia DR3 6049163679218606720 & 56903.38 & 0.04 & 0\tablenotemark{e}  & 1186983.93 & 246.6789 & -24.3416 & 1.14\\
$\cdots$ & $\cdots$ & $\cdots$ & $\cdots$ & $\cdots$ & $\cdots$ & $\cdots$ & $\cdots$\\
Gaia DR3 6046963281584426112 & 2231402.93 & 1967.07 & -10000\tablenotemark{d} & 1134.38 & 251.3130 & -24.8956 & 7.32\\
\enddata
\tablecomments{Column 1: unique source designation in \textit{Gaia}~DR3. Column 2: distance of closest approach to EM*~SR~21 within the past $10^4$~yr, i.e., the expectation of the one-dimensional normal distribution in Equation~\eqref{eq-main-parallel}. Column 3: standard deviation of the distance of closest approach, i.e., the standard deviation of the one-dimensional normal distribution in Equation~\eqref{eq-main-parallel}. Column 4: closest approach time within the past $10^4$~yr. Column 5: $z$-score, $z=d^{(f)}_{\rm c.a.}/\sigma_{d^{(f)}_{\rm c.a.}}$. This number is unitless, yet its meaning in astronomy is $\sigma$ which shows the significance from zero separation. Column 6: right ascension of the star in Column 1. Column 7: declination of the star in Column 1. Column 8: \textit{Gaia}~DR3 Renormalised Unit Weight Error (RUWE) of each source. 
\tablenotetext{a}{The table presented here is for explanation and demonstration purposes only, and thus some values are rounded to certain decimal places. See the accompanying \texttt{EM\_SR21\_sequence\_d.csv} file for a complete table for EM*~SR~21, and other \texttt{.csv} files for the other spiral hosts.}
\tablenotetext{b}{Sorted by distance of closest approach in Column 2 to EM*~SR~21 in the past $10^4$~yr.}
\tablenotetext{c}{Analytical solutions corresponding to the minimum expectation in Equation~\eqref{eq-main-parallel} from a frequentist approach.}
\tablenotetext{d}{Closest approach in the past (earlier than $10^4$~yr ago). The closest approach distance and standard deviation values are calculated for this time stamp from Equation~\eqref{eq-main-parallel}, which are not posterior distributions.}
\tablenotetext{e}{Closest approach in the future. The closest approach distance and standard deviation values are calculated for this time stamp (i.e., the \textit{Gaia}~DR3 epoch).}
}

    (The data used to create this table are available in the `anc' folder on arXiv.)

\end{deluxetable*}

\subsubsection{HD~34700 (i.e., HD~34700~A)}\label{sec-hd34700}
A large low-density cavity and multiple spiral-arm structures outside the cavity are revealed for HD~34700 using the Gemini Planet Imager in polarized scattered light \citep{monnier19}. The structures are also detected using the STIS instrument onboard the \textit{Hubble Space Telescope} in total intensity \citep{ygouf19}, and with SCExAO/CHARIS in multiple bands \citep{uyama20b}. The HD~34700 system is comprised of a binary Herbig Ae system (HD~34700AaAb) and two distant companions. Spiral fitting by \citet{uyama20b} results in large pitch angles, which might be explained by the flyby of HD~34700~B and/or HD~34700~C.

Among the 367 \textit{Gaia}~DR3 sources within $10$~pc from HD~34700 (i.e., HD~34700~A), 7 sources have their closest  on-sky approach within the past $10^4$~yr, yet 6 of them are at least $145\sigma$ away from the spiral host, except \textit{Gaia}~DR3~3237512159787015936 (i.e., HD~34700~C) which has $d^{(f)}_{\rm c.a.}=3184_{-127}^{+127}$~au (i.e., $25\sigma$ from zero distance) at $t^{(f)}_{\rm c.a.}=-3053$~yr. We do not identify other recent close-in on-sky flyby candidates.

\begin{figure*}[htb!]
	\centerfloat
	\includegraphics[width=0.95\textwidth]{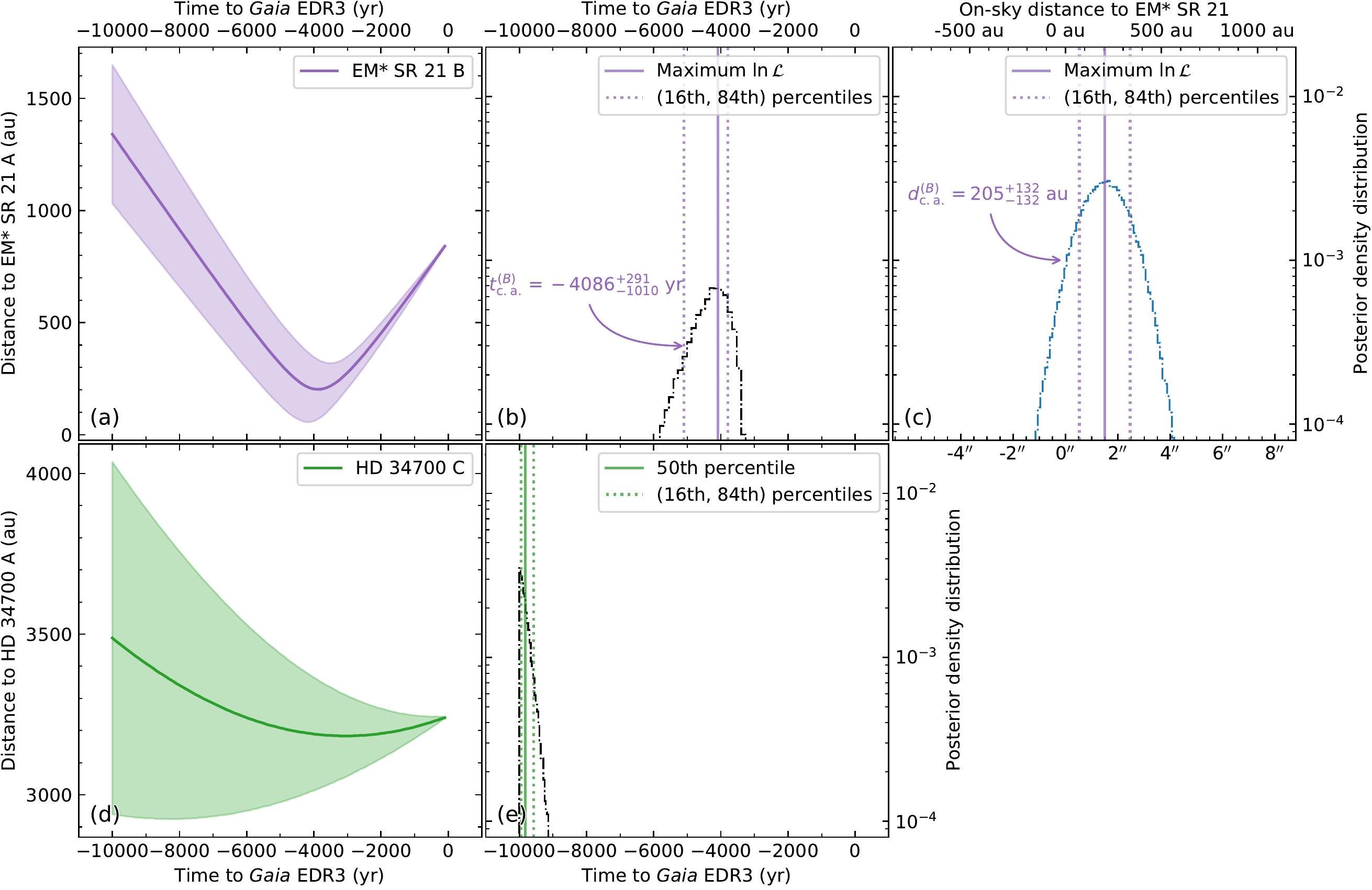}
    \caption{Relative distance as a function of past time for systems with recent close-in flyby events in Section~\ref{sec-flyby-with-credible}. Top: \textit{Gaia}~DR3~6049152203067495808 (i.e., EM*~SR~21~B) seen from EM*~SR~21: moderate uncertainty growth in \textbf{(a)} allows for posterior calculation of the time of closest approach  in \textbf{(b)}, which shares the same vertical axis with the approach distance in \textbf{(c)}. Bottom: \textit{Gaia}~DR3~3237512159787015936 (i.e., HD~34700~C) seen from HD~34700: uncertainty expansion in \textbf{(d)} cannot give conclusive answers to nearest approach time in maximum likelihood in \textbf{(e)}, in which the MCMC samples drift towards the prior limit of $-10^{4}$~yr. Note: negative distances in \textbf{(c)} are physically implausible, yet it is the framework limit when sampling from normal distributions.}
    \label{fig-dist-year}
\end{figure*}

We cannot establish a statistically well-constrained posterior flyby time for HD~34700~C within the past $10^4$ yr, see Figure~\ref{fig-dist-year}d. This is because the fast uncertainty growth of the relative distance can lead to higher likelihood values for earlier times. Nevertheless, we can identify a more past flyby at $t_{\rm c.a.}^{(B)}=-4.1_{-2.5}^{+1.4}\times10^4$~yr, which has a large uncertainty and we do not further investigate, since it is beyond the expected $10^4$~yr requirement for spirals to persist. In comparison, the \citet{uyama20b} flyby calculation for HD~34700~C is prone to systematic offsets between \citet{sterzik05} and \textit{Gaia}~DR2 \citep{gaia18}, when the latter did not include proper motion measurements for HD~34700~C. Future observations with higher precision than \textit{Gaia}~DR3 may better constrain the flyby history for HD~34700~C and resolve the difference between the frequentist and Bayesian approach.

\subsection{Isolated systems: no credible \textit{Gaia}~DR3  flyby candidates}\label{sec-flyby-no-credible}
We have not identified credible \textit{Gaia}~DR3 flyby candidates for 15 systems, see a summary of the information for all 20 systems in Table~\ref{tab:flybyinfo}.  Nevertheless, the existence of faint and close-in stars including binaries that have not been accessible from \textit{Gaia}, or those that are accessible but without motion measurements yet in \textit{Gaia}~DR3, could be the \textit{de facto} flyby candidates.

\subsection{(Un)known binary systems}\label{sec-flyby-binary}
We did not identify flybys from neighbouring stars for the following systems, but they are (or could be) binary systems which have been (or should be) confirmed. What is more, the high RUWE measurements for certain systems suggest that the \textit{Gaia}~DR3 values could be biased due to the existence of binary companions. As a result, our linear framework that is fed with biased values cannot accurately trace the historical and future locations for these stars. To finalize flyby history for spiral formation mechanisms with high RUWE values, more reliable solutions from \textit{Gaia} and a more realistic propagation framework are necessary.

\subsubsection{AS~205}
AS~205 is a binary system with the northern star AS~205~N hosting a pair of spirals in ALMA in \citet{kurtovic18}, and the southern target AS~205~S being a spectroscopic binary. Gas emission between AS~205~N and AS~205~S shows tidal interaction between them, and the spirals surrounding AS~205~N could be explained by this \citep[e.g., UX~Tau:][]{menard20}. 

Among the 793 \textit{Gaia}~DR3 sources within $10$~pc from AS~205, 6 sources have their closest on-sky approach within the past $10^4$~yr, yet they are at least $86\sigma$ away from the spiral host. AS~205~S, however, will have a future closest approach of $d^{(f)}_{\rm c.a.}=183.3_{-1.6}^{+1.6}$~au at $t^{(f)}_{\rm c.a.}=+76.3$~yr. We do not identify recent close-in on-sky flyby candidates. However, the RUWE measurement for AS~205 is 1.75, indicating non-reliable solution from \textit{Gaia}~DR3 and compromising the corresponding linear flyby results.

\subsubsection{HD~100453}
HD~100453 is a binary system with a double spiral pattern in \citet{wagner15}, with \citet{dong16} explaining the two spirals are driven by the companion HD~100453~B. \citet{rosotti20} observed CO gas connection to the companion in peak intensity using ALMA, which is consistent with the expectation that the spirals are driven by it. 

The companion HD~100453~B is not reported in \textit{Gaia}~DR3. Among the 816 \textit{Gaia}~DR3 sources within $10$~pc from HD~100453, 10 sources have their closest  on-sky approach within the past $10^4$~yr, among which \textit{Gaia}~DR3~5345034027626030848 had an on-sky distance $d^{(f)}_{\rm c.a.}=2767_{-1409}^{+1409}$~au (i.e., being $2\sigma$ away from zero separation) from the host at $t^{(f)}_{\rm c.a.}=-9740$~yr. Yet the rest of the 9 systems are at least $100\sigma$ away from the spiral host.

\subsubsection{V599~Ori}
V599~Ori shows a possible spiral in the SPHERE data. Among the 386 \textit{Gaia}~DR3 sources within $10$~pc from V599~Ori, 3 sources have their closest  on-sky approach within the past $10^4$~yr, yet they are at least $5769\sigma$ away from the spiral host. We do not identify recent close-in on-sky flyby candidates.

The SPHERE data, however, shows a visually nearby source at $2\farcs04$ to its northeast side. It matches the source \textit{Gaia}~DR3~3016081586083413120. Nevertheless, there is no parallax measurement for this source from \textit{Gaia}~DR3.

Given that \textit{Gaia}~DR3~3016081586083413120 situates along the pointing direction of the possible arm of V599~Ori, we cannot rule out it being a faint object that passed by the V599~Ori system. Future parallax and proper motion measurements is necessary to investigate the flyby relationship of the two sources.

\section{Summary}\label{sec-sum}
Recent and close-in flyby of neighboring stars can excite short-lived spiral arms in protoplanetary disks. Assuming linear motion propagation and by developing an analytical framework {\tt afm-spirals} \citep{shuai22} and performing flyby analysis for 20 known spiral arm hosts resolved in polarized scattered light with VLT/SPHERE/IRDIS, we have identified 
EM*~SR~21 as the only system with a plausible recent flyby candidate (EM*~SR~21~B) responsible for the spirals. Nevertheless, existing evidences show that the EM*~SR~21 system could be comprised of bounded objects, making the spirals instead probably resulting from companion-disk interaction.

For isolated systems that host spirals in scattered light, we have not identified recent and close-in flyby candidates from \textit{Gaia}~DR3 for them. This suggests that such flybys could not be the primary resources for the formation of spiral arms in scattered light for isolated systems. Should such objects exist, they are beyond the detection limits of \textit{Gaia}~DR3.

Moving forward, this framework can identify flyby candidates for new spiral systems, or establish flyby events when radial velocity measurements of comparable data quality are available. The flyby candidates from this framework do not require survey-level radial velocity observations, but discrete observations of individual targets. By doing so, we can reduce the observational cost in the search for stellar flyby events.   Once these relationships are confirmed, however, we note that on the one hand, we cannot exclude the possibility that the spirals are indeed not excited by stellar flyby events. On the other hand, if stellar flyby events do have occurred, the orbits of the planets in these planet-forming systems could be excited to exotic orbits \citep[e.g.,][]{nguyen21} that could be more challenging to image than those with circular orbits, which were previously assumed in spiral motion measurements \citep[e.g.,][]{ren20, xie21, safonov22}.

There are limitations in our study. First, we have only presented stellar flyby candidates that are identified in \textit{Gaia}~DR3 with both location and proper motion measurements. Second, we have adopted linear motion propagation in our flyby study, however this approach should be accurate for the purpose of identifying flyby events for isolated systems \citep[e.g.,][]{derosa19}. Third, for stars with moderate motion uncertainties -- yet low accuracy -- which thus cannot be identified to have statistically significant flyby, they could still be responsible for the excitement of spiral arms. Last but not least, the high RUWE terms in \textit{Gaia}~DR3 products for certain sources suggest biased measurements that could result into non-credible flyby events, and high RUWE values might indicate the existence of undetected companions which can drive the spirals. Nevertheless, when upcoming measurements with higher accuracy are available, our analytical framework will aid in narrowing down their connection to the observed spiral systems.

\begin{acknowledgements}
We thank the anonymous referees who provided comments and made this paper more rigorous. We thank Myriam Benisty, Gregory Herczeg, Andrew Winter, Xuesong Wang, and Bo Zhang for useful discussions. T.F.~and L.S.~are supported by the National Key R\&D Program of China No.~2017YFA0402600, and NSFC grants No.~11890692 and 12133008. T.F.~and L.S.~acknowledge the science research grants from the China Manned Space Project with No.~CMS-CSST-2021-A04. This project has received funding from the European Research Council (ERC) under the European Union's Horizon 2020 research and innovation programme (PROTOPLANETS, grant agreement No. 101002188). R.D.~acknowledges financial support provided by the Natural Sciences and Engineering Research Council of Canada through a Discovery Grant, as well as the Alfred P.~Sloan Foundation through a Sloan Research Fellowship. L.S., B.B.R., R.D., and X.Z.~acknowledge the 2022 Protoplanetary Disks and Planet Formation summer school hosted at China Center of Advanced Science and Technology (CCAST) for discussion. 

The spiral arm gallery images are based on observations collected at the European Organisation for Astronomical Research in the Southern Hemisphere under ESO programme 0104.C-0157(B) for AB~Aur, 099.C-0685(A) for AS~205, 098.C-0760(B) for CQ~Tau, 0102.C-0453(A) for DR~Tau, 097.C-0536(A) for DZ~Cha, 1100.C-0481(Q) for EM*~SR~21, 60.A-9800(S) for HD~34282, 1104.C-0415(A) for HD~34700, 096.C-0248(B)	 for HD~100453, 096.C-0248(B)	 for HD~100546, 099.C-0601(A) for HD~142527, 097.C-0902(A) for HD~143006, 098.C-0760(B) for LkHa~330, 60.A-9389(A) for MWC~758, 095.C-0273(A) for SAO~206462, 198.C-0209(F) for TWA~7,  1104.C-0415(G) for V599~Ori, 0102.C-0778(A) for V1247~Ori, 1100.C-0481(R) for Wa~Oph~6, and 098.C-0486(A) for WW~Cha. 

This work has made use of data from the European Space Agency (ESA) mission {\it Gaia} (\url{https://www.cosmos.esa.int/gaia}), processed by the {\it Gaia} Data Processing and Analysis Consortium (DPAC, \url{https://www.cosmos.esa.int/web/gaia/dpac/consortium}). Funding for the DPAC has been provided by national institutions, in particular the institutions participating in the {\it Gaia} Multilateral Agreement.
\end{acknowledgements}

\facilities{\textit{Gaia}, VLT:Melipal (SPHERE)}

\software{{\tt afm-spirals} \citep{shuai22}, {\tt emcee} \citep{emcee}}

\begin{appendix}

\section{Querying stars within $10$ ~pc}\label{sec-gaia-query-code}
For each spiral host, we query the \textit{Gaia} DR3 stars that are within its $10$~pc at \url{https://gea.esac.esa.int/archive/}, see the following Astronomical Data Query Language ({\tt ADQL}) code which uses MWC~758 as a query example. In addition to the default results from any \textit{Gaia} query, the {\tt ADQL} outputs are used for our flyby analysis by including for each neighbor its angular on-sky distance to the host (``dist\_arcsec''), distance to the Solar System (``z\_pc''), and three-dimensional physical distance to the host (``dist\_to\_host'').

\begin{lstlisting}[style=custom_adql]
SELECT DISTANCE( POINT('ICRS', ra, dec), POINT('ICRS', 82.61472046369643, 25.332403978827056))*3600 AS dist_arcsec, -- Calculate on-sky angular distance to MWC 758
1000/gaiaDR3.gaia_source.parallax as z_pc, -- Convert parallax to distance in pc from Solar System
sqrt(power(155.87, 2)+power(1000/gaiaDR3.gaia_source.parallax, 2)-2*155.87*1000/gaiaDR3.gaia_source.parallax*cos(RADIANS(DISTANCE( POINT('ICRS', ra, dec), POINT('ICRS', 82.61472046369643, 25.332403978827056))))) as dist_to_MWC758, -- Calculate 3D distance to MWC 758
* FROM gaiaDR3.gaia_source 
    WHERE 1=CONTAINS( POINT('ICRS', ra, dec), -- Select neighbours within 10 pc from MWC 758 with 3 requirements below
                    CIRCLE('ICRS', 82.61472046369643, 25.332403978827056, 4)) and -- 1. Circling 11 pc on-sky sources from MWC 758, where 4deg = arctan(11pc/155.87pc) is calculated for MWC 758
                    (1000/gaiaDR3.gaia_source.parallax < 167) and -- 2a. Within 167 pc from Solar System
                        (1000/gaiaDR3.gaia_source.parallax > 144) and -- 2b. Beyond 144 pc from Solar System
                    (sqrt(power(155.87,2)+power(1000/gaiaDR3.gaia_source.parallax,2)-2*155.87*1000/gaiaDR3.gaia_source.parallax*cos(RADIANS(DISTANCE( POINT('ICRS', ra, dec), POINT('ICRS', 82.61472046369643, 25.332403978827056))))) < 10) -- 3. 3D distance to MWC 758 within 10 pc
ORDER BY dist_to_host ASC -- Order output stars by 3D distances to MWC 758
\end{lstlisting}

In the above {\tt ADQL} query example, we increase the querying efficiency as follows. We first find the stars whose on-sky angular distances are within $11$~pc from MWC~758 to obtain a cone. We then identify those that have a distance that is within $10$~pc from the MWC~758 distance to the Solar System, thus segmenting the cone with concentric spheres. We finally select the stars that are within $10$~pc from MWC~758 in this segmented cone using the law of cosines, and order them by their three-dimensional distance to MWC~758. By doing so, the querying results can be obtained within $5$~min for each host.

\section{On-sky flyby framework}\label{sec-app-unc}

\begin{figure}[htb!]
	\centerfloat
	\includegraphics[width=0.45\textwidth]{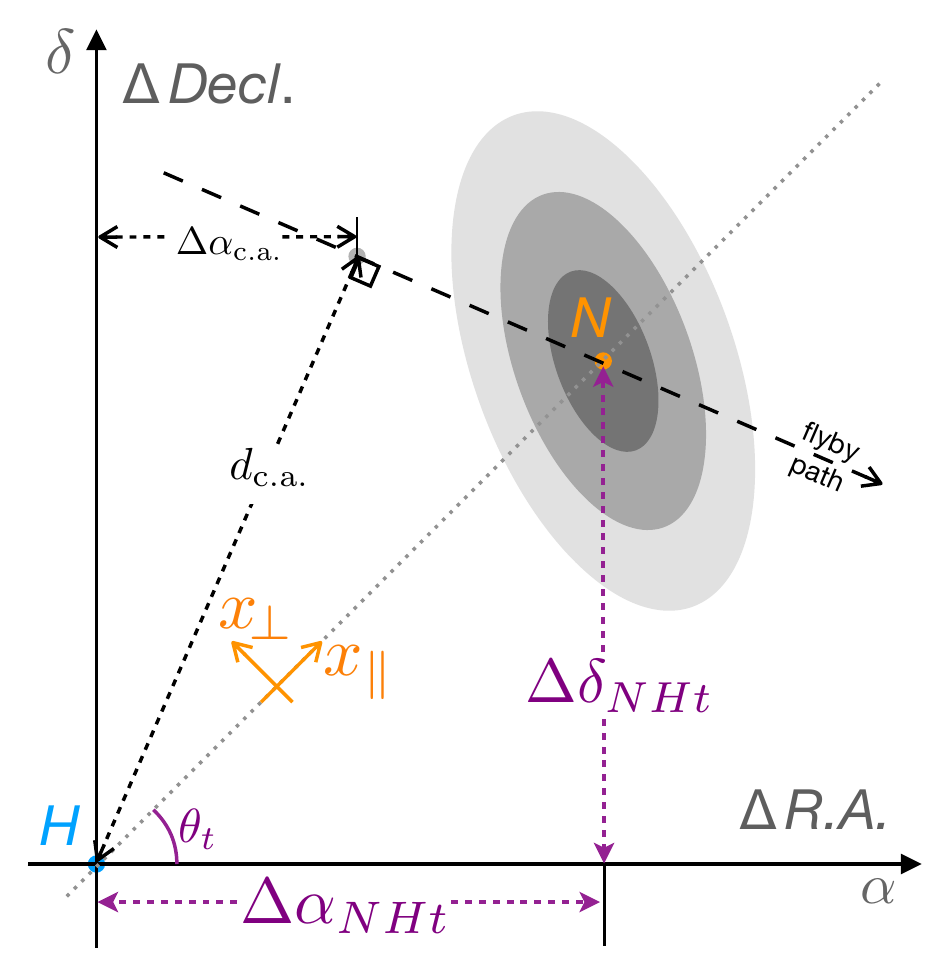}
    \caption{Sketch of bivariate normal distribution in \ref{sec-app-unc}. Point $H$ is the position of spiral host in its reference frame (i.e., origin).  Point $N$ is the relative location of a star neighbour seen from $H$, which is located at the center of the bivariate normal distribution in Equation~\eqref{eq-bivariate-normal}. The distribution along line $\overleftrightarrow{HN}$ is a one-dimensional normal distribution, see Equation~\eqref{eq-1d-normal-NHt}.}
    \label{fig-appendixdemo}
\end{figure}

\subsection{Relative position}\label{sec-app-relative-position}
The on-sky generalized coordinate (i.e., location and proper motion vector) of a star, $\bm{q} = (R.A., Decl., {\rm pm}_{R.A.}, {\rm pm}_{Decl.})^T \triangleq (\alpha, \delta, \mu_{\alpha}, \mu_{\delta})^T \in \mathbb{R}^{4\times1}$ follows a four-dimensional multivariate normal distribution
\begin{equation}
\bm{q}\sim\mathcal{N}_4(\bm{\mu}, \bm{\Sigma}),\label{eq-gen-coordinate}
\end{equation}
where $\triangleq$ is the symbol of definition, $\bm{\mu} = (\mathbb{E}\alpha, \mathbb{E}\delta, \mathbb{E}\mu_{\alpha}, \mathbb{E}\mu_{\delta})^T\in \mathbb{R}^{4\times1}$ contains the expected values from \textit{Gaia} DR3, and $\bm{\Sigma}\in \mathbb{R}^{4\times4}$ is the covariance matrix from \textit{Gaia} DR3. Specifically, we divide the default \textit{Gaia} output of $\mu_{\alpha^*}$ by $\cos{\delta}$ to obtain $\mu_{\alpha}$ \citep[e.g.,][]{Squicciarini2021}, and we modify the covariance matrix $\bm{\Sigma}$ similarly  wherever applicable.\footnote{We have experimented dividing the individual $\cos\delta$ terms and group-averaged ones, and the results have negligible difference.} 

At time $t\in\mathbb{R}$, let 
\begin{equation}
T_t = \begin{bmatrix}
1& 0 & t & 0\\ 
0 & 1 & 0 & t
\end{bmatrix}\label{eq-transf-mat}
\end{equation}
be a transformation matrix, the on-sky position vector of the star, $\bm{x}_t=(R.A., Decl.)^T\in \mathbb{R}^{2\times1}$,  is then

\begin{align*}
    \bm{x}_t=T_t \bm{q} &= \begin{bmatrix}
1& 0 & t & 0\\ 
0 & 1 & 0 & t
\end{bmatrix}\begin{pmatrix}
\alpha \\
\delta \\
\mu_{\alpha}\\
\mu_{\delta}
\end{pmatrix}\\
&=
\begin{pmatrix}
\alpha + t \mu_{\alpha}\\
\delta + t \mu_{\delta}
\end{pmatrix}.\numberthis\label{eq-rt}
\end{align*} 
{Assuming a constant $\cos{\delta}$,\footnote{{This assumption is valid for our study, since a $10$~pc radius at a typical $140$~pc from the Solar System spans ${\approx}4^\circ$ in $\delta$, which is further negligible for a $1$~pc radius at $140$~pc (i.e., $0\fdg4$).}}} $\bm{x}_t$ follows a bivariate normal distribution,  $\bm{x}_t \sim \mathcal{N}_2 (T_t \bm{\mu}, T_t \bm{\Sigma} T_t^T)$, where superscript $^T$ denotes matrix transpose. 

The relative location of star neighbour $N$ from spiral host $H$ at time $t$, $\Delta\bm{x}_{NHt}=\bm{x}_t^{(N)}-\bm{x}_t^{(H)} \triangleq (\Delta \alpha_{NHt}, \Delta \delta_{NHt})^T \in \mathbb{R}^{2\times1}$ follows a bivariate normal distribution,
\begin{align*}
\Delta\bm{x}_{NHt}\sim&\mathcal{N}_2\left(T_t \bm{\mu}^{(N)}-T_t \bm{\mu}^{(H)}, T_t \bm{\Sigma}^{(N)}T_t^T +T_t\bm{\Sigma}^{(H)} T_t^T \right) \\
			&\triangleq \mathcal{N}_2\left( \bm{\mu}_{NHt}, \bm{\Sigma}_{NHt} \right),\numberthis\label{eq-bivariate-normal}
\end{align*}
where superscripts $^{(H)}$ and $^{(N)}$ denote the generalized coordinates for spiral host $H$ and its star neighbour $N$ from Equation~\eqref{eq-gen-coordinate}, respectively.  See Figure~\ref{fig-appendixdemo} for an example of such a distribution at time $t$.

\subsection{Statistical distribution}\label{sec-app-stat-distribution}
The statistical quantification of the relative location between two stars can be obtained along the direction of the two stars, i.e., along line $\overleftrightarrow{HN}$ in Figure~\ref{fig-appendixdemo}. At time $t$, the functional form of Equation~\eqref{eq-bivariate-normal} is fixed, and we can clockwise rotate the bivariate normal distribution by angle ${\theta_t}$ to align point $N$ along the horizontal axis. The rotation matrix for such a clockwise rotation is

\begin{align*}
R_{\theta_t} =& \begin{bmatrix}
\cos{\theta_t} & \sin{\theta_t}\\ 
-\sin{\theta_t} & \cos{\theta_t}
\end{bmatrix}, \numberthis \label{eq-rot}\\
         =& \frac{1}{\sqrt{(\Delta \alpha_{NHt})^2 + (\Delta \delta_{NHt})^2}}\begin{bmatrix}
\Delta \alpha_{NHt} & \Delta \delta_{NHt}\\ 
 -\Delta \delta_{NHt} & \Delta \alpha_{NHt}
 \end{bmatrix}, \numberthis \label{eq-rot-actual}
\end{align*}
where $\Delta \alpha_{NHt} = \left(\mathbb{E}\alpha^{\left(N\right)}-\mathbb{E}\alpha^{\left(H\right)}\right) + t \left(\mathbb{E}\mu_{\alpha}^{\left(N\right)}-\mathbb{E}\mu_{\alpha}^{\left(H\right)}\right)$, and $\Delta \delta_{NHt} = \left(\mathbb{E}\delta^{\left(N\right)}-\mathbb{E}\delta^{\left(H\right)}\right) + t \left(\mathbb{E}\mu_\delta^{\left(N\right)}-\mathbb{E}\mu_\delta^{\left(H\right)}\right)$.

The corresponding expression of the bivariate normal distribution after this rotation is
\begin{widetext}
\begin{align}
\Delta\bm{x}'_{NHt}   \triangleq \begin{pmatrix}
x_\parallel\\
x_\perp
\end{pmatrix}
				\sim & \mathcal{N}_2\left( R_{\theta_t}\bm{\mu}_{NHt}, R_{\theta_t} \bm{\Sigma}_{NHt} R_{\theta_t}^T \right) \nonumber \\
                &=\mathcal{N}_2\left( R_{\theta_t} T_t \bm{\mu}^{(N)}-R_{\theta_t} T_t \bm{\mu}^{(H)}, R_{\theta_t} T_t \bm{\Sigma}^{(N)}T_t^T R_{\theta_t}^T +R_{\theta_t} T_t\bm{\Sigma}^{(H)} T_t^T R_{\theta_t}^T \right) \nonumber\\
				&\triangleq \mathcal{N}_2\left( \begin{bmatrix}
\mu_\parallel \\
\mu_\perp
\end{bmatrix} ,
\begin{bmatrix}
\sigma_\parallel^2 & \rho\sigma_\parallel\sigma_\perp\\
\rho\sigma_\parallel\sigma_\perp & \sigma_\perp^2
\end{bmatrix} 
\right), \label{eq-bivariate-normal-rotate-def}
\end{align}
\end{widetext}
where $\mu_\parallel\in \mathbb{R}^+$, and $\rho\in[-1,1]$ is the correlation. The first element of $\Delta\bm{x}'_{NHt}$, $x_\parallel$, corresponds to the distribution along line $\overleftrightarrow{HN}$, and the second element $x_\perp$ perpendicular to line $\overleftrightarrow{HN}$, see Figure~\ref{fig-appendixdemo}.

Given the information of spiral host $H$ and a star neighbour $N$, the probability density distribution along line $\overleftrightarrow{HN}$ at time $t$, is therefore
\begin{equation}\label{eq-parallel}
x_\parallel \mid t \sim \mathcal{N}_1\left(\mu_\parallel, \sigma^2_\parallel\right).
\end{equation}

\subsection{Closest approach}\label{sec-app-ca}
In the reference frame of host $H$, the expected flyby path of star neighbour $N$ is a straight line, since the proper motion measurements are linearly propagated. The minimum distance between the two stars is thus the length of the line segment that is perpendicular to the flyby path of $N$, see Figure~\ref{fig-appendixdemo}.

For $H$ and $N$, to obtain the analytical expressions of the closest approaches, here we define relative $R.A.$~offset in \textit{Gaia} DR3 measurement  as $\Delta\alpha=\alpha^{(N)}-\alpha^{(H)}$, relative $Decl.$~measurement $\Delta\delta=\delta^{(N)}-\delta^{(H)}$, relative proper motion along the $R.A.$~direction $\Delta{\mu_{\alpha}}=\mu_{\alpha}^{(N)}-\mu_{\alpha}^{(H)}$, and relative motion along the $Decl.$~direction $\Delta{\mu_{\delta}}=\mu_{\delta}^{(N)}-\mu_{\delta}^{(H)}$. Aligning the $R.A.$ direction with the $x$-axis and the $Decl.$~direction the $y$-axis in Cartesian coordinates, the flyby path can be expressed as
\[
\frac{y-\Delta\delta}{x-\Delta\alpha}=\frac{\Delta\mu_{\delta}}{\Delta\mu_{\alpha}},
\]
or
\begin{equation}\label{eq-path-flyby}
    \Delta\mu_{\delta}\cdot x - \Delta\mu_{\alpha}\cdot y +\Delta\mu_{\alpha}\cdot \Delta\delta - \Delta\mu_{\delta}\cdot \Delta\alpha = 0.
\end{equation}

Defining the functional form for the flyby path to be $ax + by + c = 0$, and comparing with Equation~\eqref{eq-path-flyby}, then we have \begin{equation}
\left\{\begin{matrix}
a = &\Delta\mu_{\delta}\\ 
b =&-\Delta\mu_{\alpha}\\ 
c =&\Delta\mu_{\alpha}\cdot \Delta\delta - \Delta\mu_{\delta}\cdot \Delta\alpha
\end{matrix}\right. .
\end{equation}
The distance of the closest approach, or the length of the line segment that is perpendicular to the flyby path from $H$, is then
\begin{equation}\label{eq-dca}
d^{(f)}_{\rm c.a.} = \frac{\left|\frac{c}{a}\right| \left|\frac{c}{b}\right|}{|c|\sqrt{\frac{1}{a^2} + \frac{1}{b^2}}} = \frac{|c|}{\sqrt{a^2+b^2}}.
\end{equation}

Focusing only on the horizontal direction, we can obtain the closest relative distance in the $R.A.$~direction,
\[  
\Delta\alpha_{\rm c.a.} = \frac{-{ac}}{{a^2}+{b^2}}.
\]

Comparing the current -- \textit{Gaia} DR3 -- relative distance in the $R.A.$~direction and the closest approach, we can obtain the time of closest approach analytically,
\begin{equation}\label{eq-tca}
t^{(f)}_{\rm c.a.} = -{\frac{\Delta\alpha-\Delta\alpha_{c.a.} }{\Delta \mu_{\alpha}}}= -{\frac{\Delta\alpha+\frac{ac}{{a^2}+{b^2}}}{\Delta \mu_{\alpha}}},
\end{equation}
where negative value corresponds to closest approach in the past, while positive value in the future.

In our analysis, we calculate both the minimum expected on-sky distance and the time of the closest approach in Equations~\eqref{eq-dca} and \eqref{eq-tca}. We compare them with our criteria in both fly-by time ($-10^4$~yr $<t_{\rm c.a.}<0$), and distance of closest approach ($0 < d_{\rm c.a.}  < 10 r_{\rm disk}$, where $r_{\rm disk}$ is the radial extent of the spirals in existing observations), to select fly-by candidates. For the selected candidates, we calculate the posterior distribution and thus credible intervals as follows.

\subsection{Posterior distribution}\label{sec-app-posterior}

For the one-dimensional distribution along line $\overleftrightarrow{HN}$ in Equation~\eqref{eq-parallel}, the corresponding probability density function is
\begin{equation}\label{eq-1d-normal-NHt}
f(x_\parallel \mid H, N, t) = \frac{1}{\sqrt{2\pi}\sigma_\parallel} \exp\left[ -\frac{(x_\parallel - \mu_\parallel)^2}{2\sigma^2_\parallel}\right].
\end{equation}

For each star pair $H$ and $N$, the \textit{Gaia} DR3 values can result into divergence  of $\tan{\theta_t}$ in the calculation of $\theta_t$ for Equation~\eqref{eq-rot}, since star $N$ can situate along the vertical axis in Figure~\ref{fig-appendixdemo} at some certain time $t$. In this study, we avoid this divergence by calculating the four matrix elements of Equation~\eqref{eq-rot-actual} based on the definition of trigonometric functions using line segments. 

\subsubsection{Closest approach time}\label{sec-app-posterior-time}
To obtain the posterior distribution for flyby times, we assume flat priors for the flyby times when $t\in[-10^4,0]$~yr. We use {\tt emcee} from \citet{emcee} to sample different flyby times $t$, and maximize the log-likelihood function corresponding to Equation~\eqref{eq-1d-normal-NHt}, i.e.,
\begin{equation}\label{eq-likelihood}
\ln \mathcal{L} (t \mid H, N) = -\frac{1}{2} \left(\frac{x_\parallel - \mu_\parallel}{\sigma_\parallel} \right)^2 - \ln \sigma_\parallel - \frac{1}{2}\ln(2\pi).
\end{equation}

For a given disk host $H$, we set $N$ to be one of its stellar neighbors and quantify the corresponding statistical significance of flyby using Equation~\eqref{eq-parallel}. We perform the likelihood calculation in Equation~\eqref{eq-likelihood} only for those candidate flyby neighbors to obtain the posterior distribution of their closest approach time $t^{(B)}_{\rm c.a.}$.

\subsubsection{Closest approach distance}\label{sec-app-posterior-distance}
For candidates with well-constrained posterior distribution of closest approach time $t^{(B)}_{\rm c.a.}$, we obtain the posterior distribution of closest approach distance $d^{(B)}_{\rm c.a.}$ using both Glivenko--Cantelli theorem \citep[e.g.,][]{chung2001book} and probability integral transform \citep{rosenblatt1952} as a proxy \citep[e.g., Appendix B of][]{ren19}.

1. Obtain the empirical cumulative distribution function (ECDF) for the posterior distribution of the closest approach time $t^{(B)}_{\rm c.a.}$.

2. Sample a uniform distribution from 0 to 1, and match it to the ECDF value of $t^{(B)}_{\rm c.a.}$ to sample different closest approach time. The corresponding sample $t$ then follows the posterior distribution of closest approach time.

3. For each sample of $t$ from the posterior distribution, obtain the analytical expression of the host-neighbor separation from Equation~\eqref{eq-parallel}. Obtain one separation sample $d$ from the one-dimensional normal distribution.

4. Repeat Steps 2 and 3 for $10^4$ times, and obtain the posterior distribution for the closest approach distance  $d^{(B)}_{\rm c.a.}$.

\subsection{Generalizing framework for future measurements}

For studies involving radial location and radial velocity from the Sun, a generalization of the framework here can be obtained by expanding the generalized coordinate $\bm{q}$ to a 6-element vector, adopting a $3\times4$ transformation matrix in Equation~\eqref{eq-transf-mat}, and using a 3 dimensional rotation matrix in Equation~\eqref{eq-rot} such as in \citet{gagne18}. 

We do not generalize the framework here since the \textit{Gaia} DR3 data do not have such complete information, and we leave it for a future study.

\end{appendix}
\bibliography{refs}
\end{CJK*}
\end{document}